# A Wrist-Worn Multimodal Reaction Time Monitoring Device for Ecologically-Valid Cognitive Assessment

Abhigyan Sarkar, Boris Rubinsky

Department of Mechanical Engineering

University of California Berkeley, Berkeley CA 94720

## Abstract

Reaction time (RT) is a fundamental measure in cognitive and neurophysiological assessment, yet most existing RT systems require active user engagement and controlled environments, limiting their use in real-world settings. This paper introduces a low cost wrist-worn instrumentation platform designed to capture human reaction times (RT) across auditory, visual, and haptic modalities with millisecond  latency in real-world conditions. The device integrates synchronized stimulus delivery and event detection within a compact microcontroller-based system, eliminating the need for user focus or examiner supervision. Emphasizing measurement fidelity, we detail the hardware architecture, timing control algorithms, and calibration methodology used to ensure consistent latency handling across modalities. A proof-of-concept study with six adult participants compares this system against a benchmark computer-based RT tool across five experimental conditions. The results confirm that the device achieves statistically comparable RT measurements with strong modality consistency, supporting its potential as a novel tool for non-obtrusive cognitive monitoring. Contributions include a validated design for time-critical behavioral measurement and a demonstration of its robustness in unconstrained, ambient-noise environments. It offers a powerful new tool for continuous, real-world cognitive monitoring and has significant potential for both research and clinical applications.

1. Introductions

In this paper, we present the design and performance evaluation of a novel, unobtrusive device capable of measuring reaction time under ecologically valid conditions, while the user engages in everyday activities. The device responds to haptic, auditory, and visual stimuli and represents the first system of its kind. Owing to its low cost and versatility, it has the potential to become a widely adopted tool in research fields that investigate human reaction time. Human reaction time (RT), the interval between stimulus onset and the initiation of a behavioral response, is a fundamental neurophysiological metric that reflects the integrated function of sensory perception, neural transmission, cognitive processing, and motor execution. As a noninvasive and sensitive indicator of sensorimotor and cognitive status, RT has been extensively employed across disciplines including cognitive neuroscience, clinical diagnostics, sports science, and

human–machine interface design. RT tasks are typically classified by complexity: *simple* RT involves a single stimulus-response pairing; *choice* RT requires selecting among multiple responses; and *discrimination* RT tasks demand response inhibition to irrelevant stimuli. (Fig 1) These paradigms yield insight into attention, memory, executive control, and processing speed. RT is influenced by factors such as age, fatigue, arousal, stimulus modality (e.g., visual, auditory, haptic), and underlying neurological or psychiatric conditions. Clinically, RT has been used to detect early cognitive decline in Alzheimer's and Parkinson's diseases [1] [2] [3], evaluate postoperative cognitive function [4], and monitor recovery from stroke or traumatic brain injury [5]. It also serves to quantify the pharmacodynamic effects of medications [6], assess cognitive impacts of fatigue and sleep deprivation [7] and support neurodevelopmental evaluations in disorders such as ADHD [8]. In sports science, RT measurements inform neuromuscular readiness and reaction training [9]. while in education, they have been used to monitor student engagement and optimize learning strategies [9]. This overview highlights the breadth of RT applications as a valuable tool for understanding and tracking human cognitive and motor function.

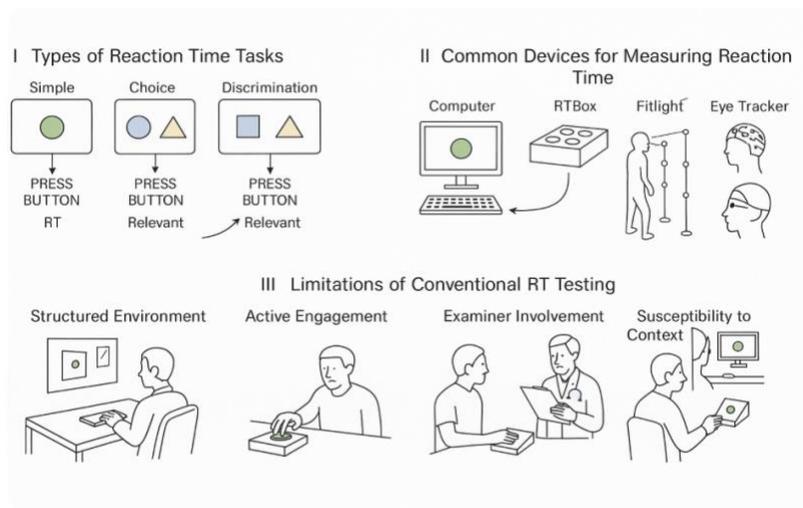

Fig. 1. Overview of Reaction Time Tasks, Measurement Devices, and Limitations of Conventional Testing  Panel I illustrates the primary categories of reaction time (RT) tasks: simple, choice, and discrimination, each requiring the user to respond to visual stimuli by pressing a button. Panel II shows commonly used devices for RT measurement, including computer-based systems, RTBox, FitLight trainers, and eye-tracking systems. Panel III summarizes key limitations of traditional RT testing, such as the need for structured environments, active participant engagement, examiner supervision, and susceptibility to contextual interference.

Despite the diversity of technologies developed to measure reaction time (RT), most systems, whether mechanical, laboratory-based, wearable, or web-based, require the user's full attention and active engagement within structured environments. From early devices like the Hipp chronoscope used by Donders [10], [11] [12] to modern tools such as E-Prime [13] Inquisit [14] and RTBox  [15], RT assessment has largely relied on stimulus–response paradigms administered under supervision or in experimental settings. While these systems offer high temporal precision and enable complex tasks (e.g., Stroop, Go/No-Go, Simon), they are inherently artificial and sensitive to context effects, such as performance anxiety or altered arousal due to observation. Even contemporary tools for neuromuscular training (e.g., Fitlight Trainer™, BlazePod™ [16]) saccadic RT assessment (e.g., Tobii Pro [17] EyeLink [18]) and neural signal analysis using EEG and ERPs [19], share this limitation. Web-based platforms like the Human Benchmark RT Test [20] and Cambridge Brain Sciences [21] have improved scalability and accessibility, but they too depend on user focus, intentional task execution, and controlled stimulus presentation. Although some systems now support multisensory input and contextual metadata capture [22] most current RT technologies remain task-focused and require user compliance. This underscores the unmet need for unobtrusive, real-time RT monitoring solutions capable of passively assessing cognitive responsiveness during unstructured, everyday activities (Fig 1).

Our group's earlier work (Ivorra et al. [23]) addressed the unmet need for a reaction time (RT) assessment technology that is both minimally obtrusive and suitable for continuous use during routine daily activities. The pioneering goal of that study was to design and validate a wrist-worn system that could deliver RT tests without requiring user focus or participation in a structured test environment. Replacing conventional responses such as button presses with a natural wrist rotation, the device enabled RT assessments to be conducted passively and context-independently. Technically, the compact 29 g device (55 × 35 × 15 mm) integrated a PIC16F689 microcontroller, dual ADXL202 accelerometers for motion detection, a vibration motor for haptic stimulation, EEPROM for data logging, and a step-up voltage regulator, all powered by a 1.2 V NiMH battery. It administered RT tests using an 800 ms preparatory and 65 ms reaction vibration stimulus, with wrist-turn responses detected via 122 Hz differential acceleration sampling. In an 8-hour real-world trial involving ten participants and 33 randomized interrogations, the system demonstrated >95% valid response rates and minimal interference with daily life, validating the feasibility of real-time cognitive monitoring in naturalistic settings. Building on this foundational work, [24], [25] advanced the field of unobtrusive RT monitoring by further refining wearable systems for continuous cognitive assessment. While Ivorra et al. [23] focused on engineering feasibility and behavioral integration, Cinaz's et al., [24], [25] studies emphasized psychometric validation and performance benchmarking. In their 2011 study, [24], Cinaz et al. introduced a wrist-mounted device combining a haptic stimulus with an inertial measurement unit (IMU), and systematically compared its RT measurements to those from conventional desktop systems under varying cognitive loads. The results confirmed comparable RT values and showed reduced perceived mental workload in the wearable, dual-task condition. Their 2012 follow-up [25] further extended the system's capability enabling richer interpretation of movement in mobile settings. Together, these studies represent a clear trajectory from Ivorra et al.'s proof-of-concept demonstration of feasibility in everyday life toward more comprehensive experimental validation, reinforcing the viability of wearable RT technologies for unobtrusive, real-world cognitive monitoring.

Accurate measurement of reaction time (RT) critically depends on minimizing and accounting for device latency, the total system delay between stimulus delivery and the reliable detection of a user response. Even delays as small as a few milliseconds can significantly bias RT data, especially in cognitive and clinical assessments where differences of 10–50 ms are diagnostically meaningful. Standard input devices such as keyboards, touchscreens, and mice often introduce unpredictable latency due to hardware polling rates, buffer delays, and asynchronous operating system processes interrupts [26]. These limitations compromise timing precision and undermine the reliability of RT measurements. This variability in performance across systems was further highlighted by Cagnotto et al. [27] who found substantial differences in accuracy and precision among commonly used RT tools, emphasizing the importance of validated instrumentation in both experimental and applied research contexts. To address this, dedicated platforms such as RTBox [15] and E-Prime response boxes [13] have been developed to provide calibrated timing circuits and synchronized stimulus-response logging. While highly accurate, these systems are tethered to structured environments and restrict mobility. In contrast, microcontroller-based systems, such as those built using Arduino, have demonstrated that when properly designed and calibrated, portable RT systems can achieve high timing accuracy with minimal latency, offering a promising path for mobile and wearable applications [28].

Wearable systems using natural movement modalities such as wrist rotation offer key advantages over conventional response mechanisms like button pressing. Button-based input introduces mechanical variability, such as actuation delay and inconsistent force thresholds, which can distort true RT values [29]. In contrast, the use of wrist rotation as a response modality introduces a biomechanically natural and unobtrusive alternative, particularly suited for wearable systems. Unlike button pressing, which requires specific posture, visual attention, and conscious engagement, wrist movement can be performed reflexively and seamlessly during everyday activities. It allows for more ecological deployment of RT tasks without disrupting natural behavior with high temporal fidelity. Additionally, wrist movement avoids the mechanical variability associated with button travel distance, force thresholds, and tactile feedback inconsistencies, all of which introduce additional temporal noise in RT measurements, e.g. [29].

Subsequent advances in wearable haptics and motion tracking have further reinforced the viability of wrist-based interaction for RT assessment. Wearable haptic displays [30] [31], and force-feedback systems[32], [33] have shown that the wrist is a suitable and sensitive site for both stimulation and response capture. These systems not only provide real-time motion data but can also deliver modulated haptic cues for stimulus delivery. For example, Adeyemi et al. [32] developed custom voice-coil actuators for precise haptic feedback at the wrist, while Sarac et al. [31] demonstrated differential perception of normal and shear skin stimuli via wearable haptic bracelets. Such advances support the shift toward wrist-centric, multisensory RT systems capable of operating in dynamic, real-world environments. Furthermore, the modularity and adaptability of modern haptic and sensing technologies are expanding the capabilities of RT monitoring in extended reality (XR) and mobile health applications [34] [35]. These innovations pave the way for continuous, context-aware RT assessment that aligns with natural behavior, enhancing usability, reducing cognitive load, and enabling broader deployment in domains such as neurocognitive screening, fatigue detection, and immersive human–machine interaction.

Collectively, these considerations support the growing shift toward wearable RT technologies that replace conventional button-based interfaces with motion-based alternatives, offering enhanced ecological validity, reduced measurement error, and greater user compliance in naturalistic settings.

The study of reaction times to auditory, haptic (touch), and visual stimuli is a common research area in psychology and neuroscience, e.g. [36], [37], [38], [39]. This research helps to understand how the brain processes information from different senses and how quickly individuals respond. Recently, Yoshida et al. [40], [41] pioneered the development of a mobile platform for investigating human reaction times (RTs) using a native iOS application developed in Swift v5.7.1 with Xcode v14.1. The system delivered visual, auditory, and haptic stimuli, individually and in combination, on an Apple iPhone 11. Their system presented stimuli at random intervals using the phone's screen flash (visual), 500 Hz tone (auditory), and Core Haptics Framework (Apple Inc.) (haptic), with users responding via manual touchscreen press. While their results confirmed that tri-modal stimuli significantly improve RT speed and consistency, particularly under cognitive load, their system, like nearly all conventional RT platforms, still requires user focus, visual attention, and deliberate interaction with the device. These limitations constrain its utility in naturalistic, continuous, or passive monitoring scenarios.

To address these critical limitations, we developed a novel wrist-worn reaction time (RT) platform that eliminates the need for screen interaction or structured task engagement. The system is minimally obtrusive, delivering brief haptic, auditory and visual cues and detecting RT through natural wrist rotation using a three-axis gyroscope. Unlike previous systems that depend on deliberate button presses or visual fixation, our device enables RT measurement during real-life activities, such as walking, conversing, or performing daily tasks, without interrupting behavior or requiring supervision. This marks a paradigm shift in RT monitoring, supporting real-time, ecologically valid cognitive assessment suitable for applications in fatigue detection, cognitive decline evaluation, and ambulatory neuropsychological screening. The key innovation lies not only in the sensing modality but in the system's capacity for passive, continuous integration into everyday life, representing a significant advancement over smartphone-based RT tools. By combining multimodal stimulus delivery, naturalistic usability, and latency-calibrated performance, the device addresses a long-standing gap in the field and offers a powerful new tool for cognitive monitoring in both research and clinical contexts.

2. Materials and Methods

2.1 Study Design and Overview

This paper introduces a minimally obtrusive, multimodal wearable device designed as an alternative to conventional reaction time (RT) assessments, which typically require individuals to sit in front of a computer and respond to stimuli in structured, artificial environments. In contrast, the proposed wrist-worn device captures and records RT to auditory, visual, and haptic stimuli in real-world settings. Stimuli are delivered through a programmable input system that enables protocol-based scheduling and modality control, allowing for consistent stimulus presentation without requiring focused attention on the device or supervision by an examiner.

The primary objective of this paper is to describe the design and implementation of the wearable system to facilitate its replication and further development by other researchers. To illustrate its performance and demonstrate feasibility, we conducted a limited validation study involving six participants. The device's accuracy and reliability across sensory modalities were compared to those of a conventional computerized RT test. Importantly, the tests were administered while participants engaged in natural conversations, simulating real-life conditions. The comparison focused on response accuracy and modality-specific latency to assess the device's validity and practical utility as a substitute for traditional, screen-based RT evaluations.

## 2.2 Wearable Device
### 2.2.1 Hardware

The wearable device utilizes an Arduino Nano 33 BLE Sense Rev2 (Arduino, Ireva, Italy), which is built upon the Nordic nRF52840 microcontroller, to perform reaction time measurements and store the collected data. It is powered by a 2S 450 mAh Lithium Polymer (LiPo) battery (Shenzhen SaiEnfeng Technology Company, Shenzhen, China) and includes two peripheral output components: (1) a mini 12000 RPM DC 3V Vibration Motor (Tatoko Shop, Xingning, China) to deliver haptic stimulus and (2) a DC 3V Active Buzzer (Bnafes, Shenzhen, China) to deliver auditory stimulus. Visual stimulus is delivered through the Arduino's onboard LED.

The Vibration Motor draws approximately 85 mA of current [42], which exceeds the Arduino Pins' current limits (15 mA per pin; 50 mA total from the 3.3 V rail) [43] [44]. Therefore, the vibration motor is powered from the battery and wired in parallel to the Arduino. Since a 2-cell LiPo battery outputs a nominal voltage of 7.4 V (ranging from 8.4 V fully charged to ~6.0 V when depleted) [45], a DSN-360-MINI DC-DC buck converter (Shenzhen Aisidesi Technology Company, Shenzhen, China) steps the voltage down to 3.3 V and protects the motor, which is rated for 1.5–3.7 V. The motor is switched by a 2N3904 NPN Transistor (BOJACK Electronics, Jieyang, China), controlled by the Arduino's analog A0 pin through a 1 kΩ base resistor (Chanzon Technology, Shenzhen, China). A 1N5819 Schottky Diode (Chanzon Technology, Shenzhen, China) is also wired in parallel with the motor as a flyback diode to protect against voltage spikes from inductive flyback. Additionally, wiring a 0.1 µF capacitor in parallel with the diode and vibration motor should be explored to filter motor noise, but no significant disruptive noise was encountered in testing. The 3V Active Buzzer draws around 25 mA of current [46], which falls within the Arduino +3V3 pin's supply capacity [44] and is therefore powered directly from it. It is switched by a 2N3904 NPN Transistor connected to the Arduino's digital pin (D2) and doesn't require a flyback diode as it contains no inductive components.

All of the device's components are wired with 22 AWG solid core wire (Adafruit, Industrial City, United States) while the battery is wired to the Arduino and the Buck Converter with a 20 AWG multi-core JST plug connector (CHEN GU, Dongguan, China). The device's complete electronic design is shown in Figure 3 (circuit schematic) and Figure 4 (labeled build).

The Arduino, Vibration Motor, and Active Buzzer are mounted onto a 2 cm wide strip of 0.8 mm thick polycarbonate plastic, which is inserted between the layers of a 2-layer cotton athletic wristband (Couver Corporation, Norwalk, United States). The polycarbonate plastic acts as a rigid mount for the components, and a cotton wristband is used due to cotton's superior static resistance compared to other fabrics (Figure 2) [47]. The wristband has three slots cut into it: (1) a 10 cm front slot for inserting the electronics which is resealed with hot glue, (2) a 0.5 cm side slot for the battery cable, and (3) a 2 cm rear slot to expose Arduino's Micro USB port and LED. The LED is also covered in a hemisphere of hot glue to diffuse the light.

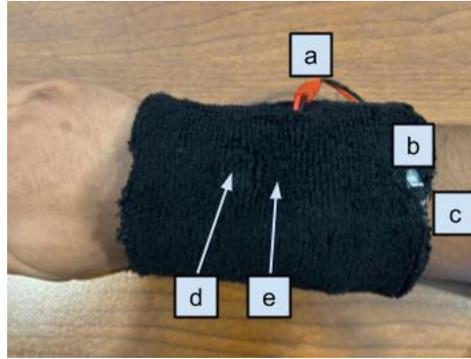

Fig. 2. Fully assembled wearable device. (a) Battery cable connected to the battery positioned beneath the hand. (b) Arduino LED used for visual stimulus delivery, enclosed in a hemispherical hot glue dome for light diffusion and protection. (c) Arduino Micro USB port for establishing a serial connection to a host system. (d) Buzzer (located beneath the fabric and not visible) used for auditory stimulus delivery. (e) Vibration motor (also beneath the fabric and not visible) used for haptic stimulus delivery .

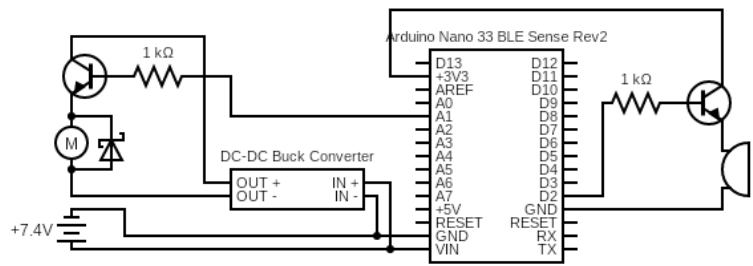

Fig. 3. Schematic diagram of the stimulus delivery and control circuitry. The system is powered by a 7.4 V battery regulated by a DC-DC buck converter, supplying power to the Arduino Nano 33 BLE Sense Rev2 microcontroller. The vibration motor is controlled via a 2N3904 NPN transistor, with a 1 kΩ base resistor and a flyback Schottky diode for inductive flyback protection. The buzzer is similarly driven by a second NPN transistor with a 1 kΩ base resistor. GPIO pins from the Arduino control the base of each transistor to activate the respective stimuli.

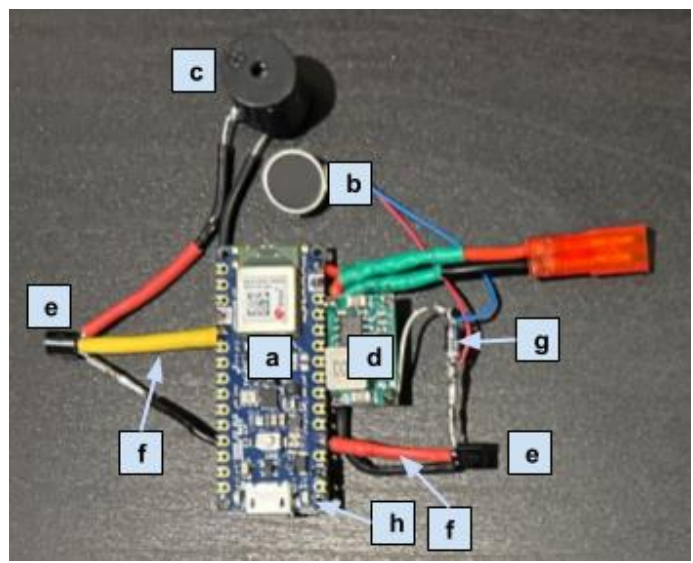

Fig. 4. Labeled components of the device's physical build. (a) Arduino Nano 33 BLE Sense microcontroller. (b) 12,000 RPM, 3 V DC vibration motor. (c) 3 V DC active buzzer. (d) DSN-360-MINI DC-DC buck converter. (e) 2N3904 NPN transistor for motor control. (f) 1 kΩ base resistor. (g) 1N5819 Schottky diode for inductive flyback protection. (h) Arduino onboard LED for visual stimulus delivery.

## 2.2.2 - Device Programming/Logic

The device is programmed in C++ using the Arduino IDE. The examiner provides commands to the device in a predetermined pattern in the form of an array of stimuli and corresponding times in seconds since the start of the test to administer the stimuli. The three types of stimuli are Auditory, Visual, and Haptic. The device has two states: (1) an experiment state where the device administers stimuli and records RT (reaction time) to the Arduino's flash storage, and (2) a read-only state where the device communicates recorded data through a serial connection to a computer. Flash storage and serial communication were selected due to their simplicity and sufficient capacity for the low amount of data collected in the experiment. This modality is sufficient for an experimental setting but requires preset stimuli commands and post-experiment serial downloading; Bluetooth should be considered for further iterations to enable real-time data transmission and remote stimulus commands.

The device automatically selects a state post-initialization; if it detects a serial connection, it will switch to data-read mode, or otherwise will begin the experiment. The device enters initialization mode as soon as power is supplied, and executes this sequence: (1) Read for serial connection, (2) Initialize IMU and sound the buzzer if IMU initialization fails. Once initialized, the device follows the procedure detailed below and in Figure 5.

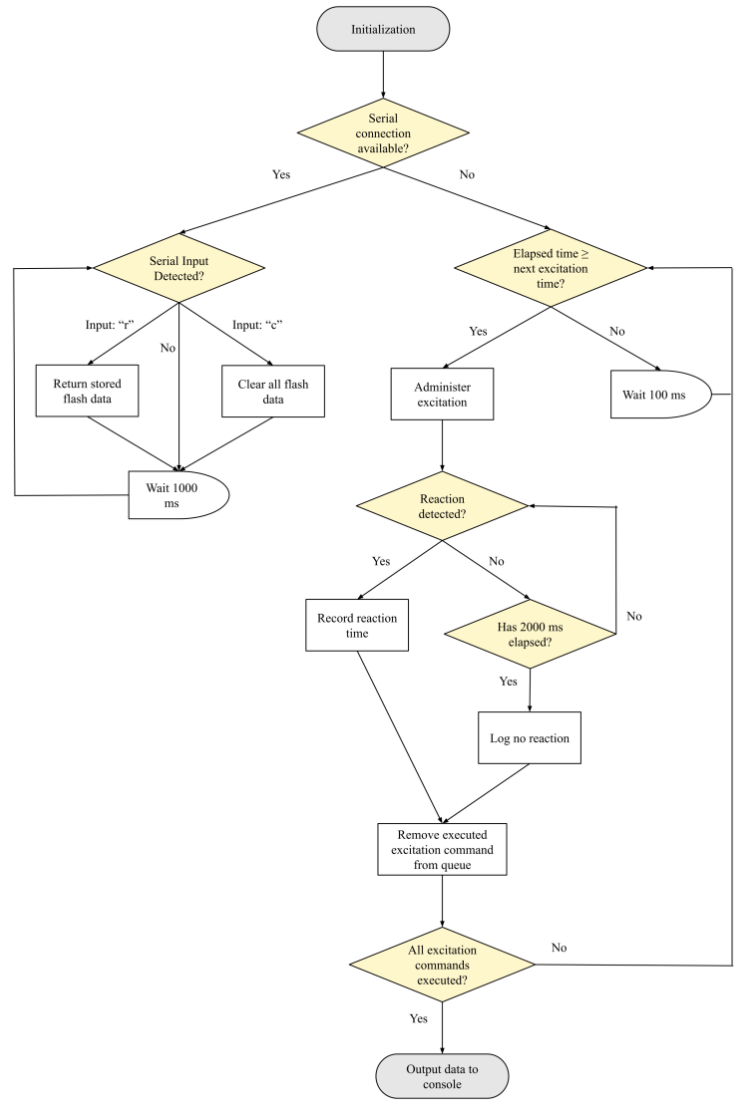

Fig. 5. Flowchart of device-side serial communication, stimulus scheduling, and reaction logging logic.

In the Serial read-only state, the device checks the serial connection for input every 1000 ms. If the input is "r", the device returns all stored data from flash, and if the input is "c", the device clears its flash memory.

In the experiment state, the device searches the predetermined queue of stimulus commands and executes the earliest command when its scheduled time has elapsed. The corresponding stimulus is delivered by setting the I/O pin responsible for controlling a certain stimulus to its HIGH state (3.3V). The stimulus remains active while the device polls for a reaction using inbuilt LSM9DS1 IMU sensor's three-axis gyroscope. A reaction is defined as a fast rotation of the wrist to the left or right which exceeds an angular velocity of ±400°/s on the x axis (i.e., the axis running horizontally across the width of the wrist). If such a gesture is measured, the device logs the reaction time as the time elapsed between the stimulus onset and gesture in milliseconds. If no reaction is detected in 2000 ms, a value of -1 is logged to indicate no reaction.

Following a reaction or timeout, the stimulus command is purged from the queue and the device returns to polling for the next stimulus command. When all commands have been executed, the device writes the collected data to its onboard 1 MB flash memory. Upon a successful data write, the device signals completion by alternating the buzzer and vibration motor twice in a predefined 1-second interval sequence. If there is a flash writing error, an error sequence is executed: the LED turns on, followed by a rapid sequence of vibrations and a final LED off state. After either sequence, it is safe to unplug the device.

### 2.2.3 - Computer Test Programming

The computer test was designed to mirror the logic of the wearable device but is limited to measuring auditory and visual reaction time, as standard computing hardware does not have the means for administering haptic stimuli. The test is programmed as a web application using JavaScript and HTML and administered on a 13" 2020 M1 Macbook Pro. JavaScript was selected due to its low latency in measuring input and outputting stimuli. An initial implementation was programmed in Python, but was discarded due to high latency with the tkinter module.

Each stimulus command is defined to the computer as a pair consisting of onset time and stimulus type. The computer test is functionally similar to the device's experiment state and follows the loop detailed below and in Figure 6.

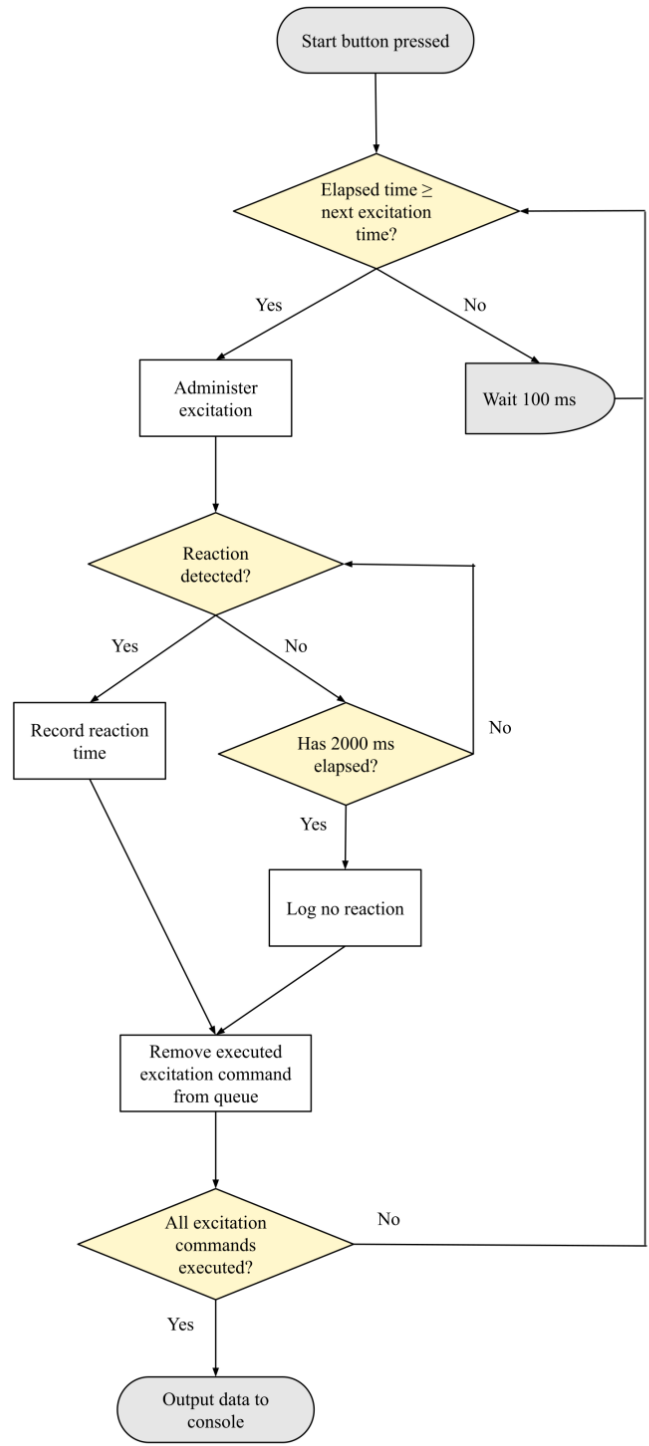

Fig. 6. Flowchart of computer-side stimulus scheduling and reaction logging logic.

The test begins when the "Start" button is pressed. Similar to the device, the researcher creates a predetermined queue of stimulus commands prior to the test. The program searches the queue to check whether the scheduled time of the earliest command has elapsed. If it has, the device administers the commanded stimulus: an auditory command triggers a "beep" sound through the computer's speaker, while a visual command changes the screen color to green (see Figure 6).

Following stimulus presentation, the program enters a 2000 ms polling window during which it monitors for a spacebar press as the user's reaction. If the spacebar is clicked in the 2000 ms interval, the program logs the time elapsed in milliseconds as reaction time. If the stimulus times out, a -1 is logged. Following a response or timeout, the respective command is purged from the queue and the program starts polling for the next command. Once all commands are executed, the program outputs all the reaction times to the console.

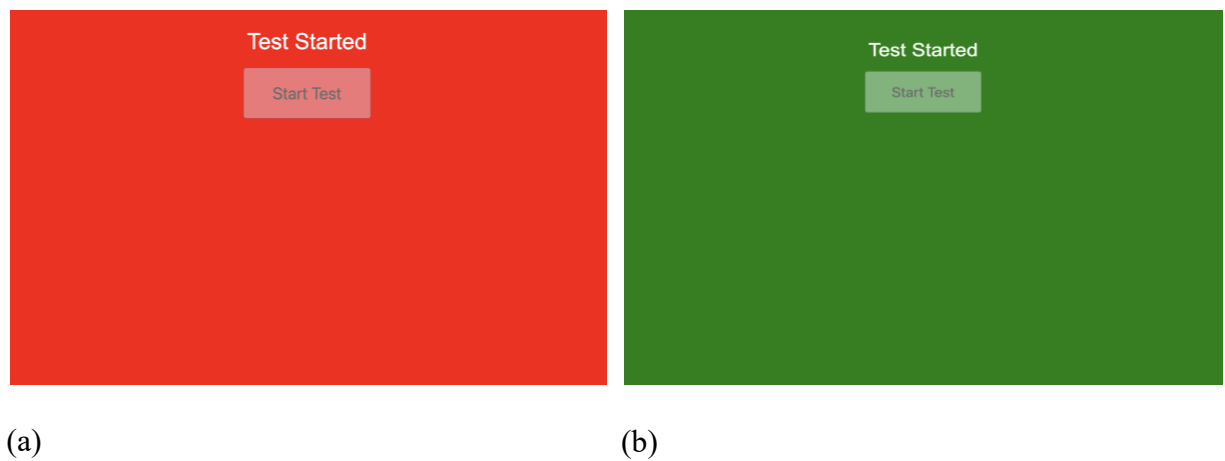

(a)                                                                 (b)

Fig. 7. Visual Computerized RT Test States. (a) The red, pre-stimulus state. (b) The green, stimulus state

### 2.2.4 - Experiment Procedures

The experimental protocol was designed to evaluate the wearable device by comparing its measurements of simple reaction time (SRT) to those obtained using a traditional computer-based test. Six healthy participants were recruited for the study, and informed consent was obtained prior to participation. The study included five sub-tests, each consisting of ten stimuli of a single modality: computer auditory, device auditory, computer visual, device visual, and device haptic. Stimuli were presented at randomized intervals ranging from 30 to 45 seconds, with timings generated using the Python random module to ensure unpredictability.

To replicate the naturalistic context in which the device is intended to operate, participants were encouraged to maintain a casual conversation with the experimenter throughout the testing session. At the beginning of each session, participants were fitted with the device on the dorsum of the non-dominant hand, secured with a rubber band to ensure firm skin contact, and asked to place the instrumented hand on a table (Figures 8 and 9).

The testing sequence proceeded as follows: (1) a brief 10-second auditory acclimation test was administered using the computer to familiarize participants with the task; (2) the computer auditory test was conducted using 10 stimulus instances at randomized times; (3) after a one-minute rest, participants underwent a 10-second auditory acclimation with the device, followed by (4) the auditory device test; (5) a 10-second visual acclimation test was then presented via the computer, followed by (6) the computer visual test (Figure 8); (7) this was followed by a 10-second device-based visual acclimation and (8) the device visual test; (9) finally, participants received a 10-second haptic acclimation and (10) the device haptic test. The same randomizing program was utilized for all stimulus types. At the conclusion of the session, participants were debriefed and provided with a summary of their average reaction times across all five test conditions.

The participants were debriefed after the study and all stated that the experiment was not stressful and the device was comfortable to wear. Additionally, participants found the test and response gesture intuitive and easy to operate. These qualitative observations support the device's applicability in longer duration natural assessments.

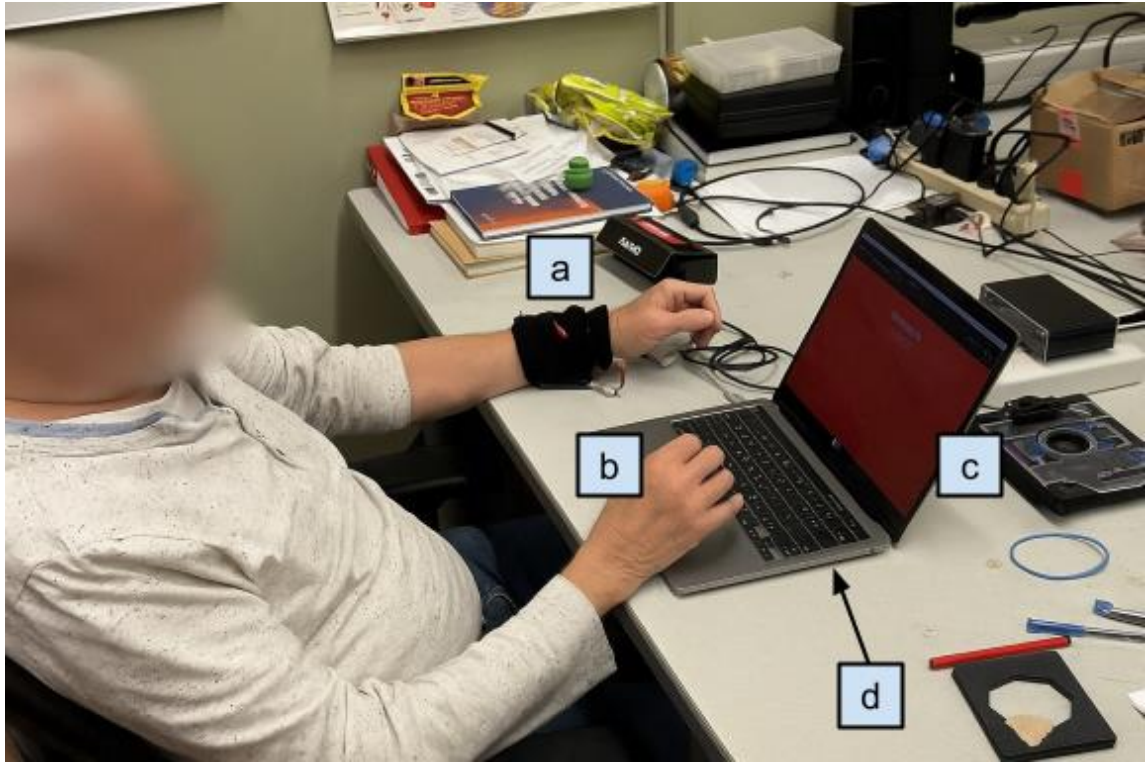

Fig. 8. Participant setup during the computer-based visual reaction time test. (a) The wearable device is not in use; the participant's non-dominant arm is resting. (b) The participant's dominant hand is positioned on the keyboard, prepared to press the spacebar. (c) The computer screen in its pre-stimulus state, ready to deliver visual stimuli. (d) The computer's speakers, used to deliver auditory stimuli.

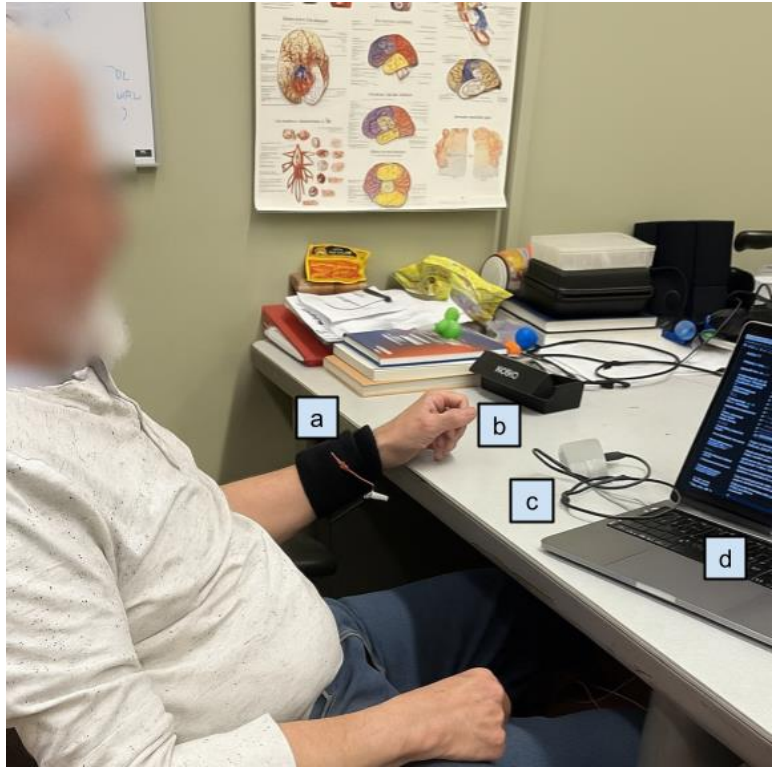

Fig. 9. Participant wearing the reaction time monitoring device during testing. (a) The device secured to the participant's non-dominant hand using a rubber band. (b) The hand resting on the table, as required by the testing protocol. (c) A Micro USB cable used to establish serial communication between the device and the examiner's computer. (d) The computer used to collect and download participant data.

### 3. Results and Discussion

The primary objective of this study was to introduce the design and implementation of a wrist-worn, multimodal reaction time (RT) monitoring device, with the aim of enabling replication and further development by other researchers. The total device costs around $55 to make. The main expensive components are the Arduino, battery, voltage converter, and wristband. The transistors, wiring, motor, flyback diode, and buzzer are near negligible in cost, with them all costing less than $1. The device takes around 2-3 hours to assemble with our design. This should make the technology widely accessible to researchers in cognition and neurobiology.

To evaluate the device's performance and demonstrate its feasibility, we conducted a limited validation study involving six adult participants. The device's accuracy and reliability across three sensory modalities, auditory, visual, and haptic, were compared against those of a conventional computerized RT assessment. Importantly, testing was performed while participants engaged in casual conversation, simulating naturalistic, real-world conditions. The comparison focused on response accuracy and modality-specific latency to assess the wearable system's validity and practical utility as an alternative to traditional, screen-based RT evaluation methods.

Participants (ages 27–77) completed trials under five experimental conditions: computer-auditory, device-auditory, computer-visual, device-visual, and device-haptic. The study aimed to validate the temporal fidelity of the wearable system, characterize variability across sensory modalities, and evaluate its applicability for real-world cognitive monitoring.

Crucially, this investigation serves as a proof of concept rather than a population-level analysis of RT performance. The small sample size was selected to enable detailed, participant-level comparisons and to demonstrate the device's feasibility and consistency across sensory modalities. Although not statistically powered for broad generalization, the study offers compelling preliminary evidence that the wearable system yields reaction time measurements comparable to those obtained from established computer-based assessments. These findings support the rationale for expanded validation in larger and more diverse populations.

The device is specifically designed for use during routine daily activities. To reflect this intended application, participants were encouraged to engage in natural conversation throughout the testing sessions. As a result, they were required to distinguish task-relevant stimuli from ongoing verbal interaction. Accordingly, the measured responses are better characterized as discrimination reaction times rather than simple reaction times, a distinction that should be considered when interpreting the findings and in guiding future applications of the device. A summary of the experimental results is presented in Figure 10.

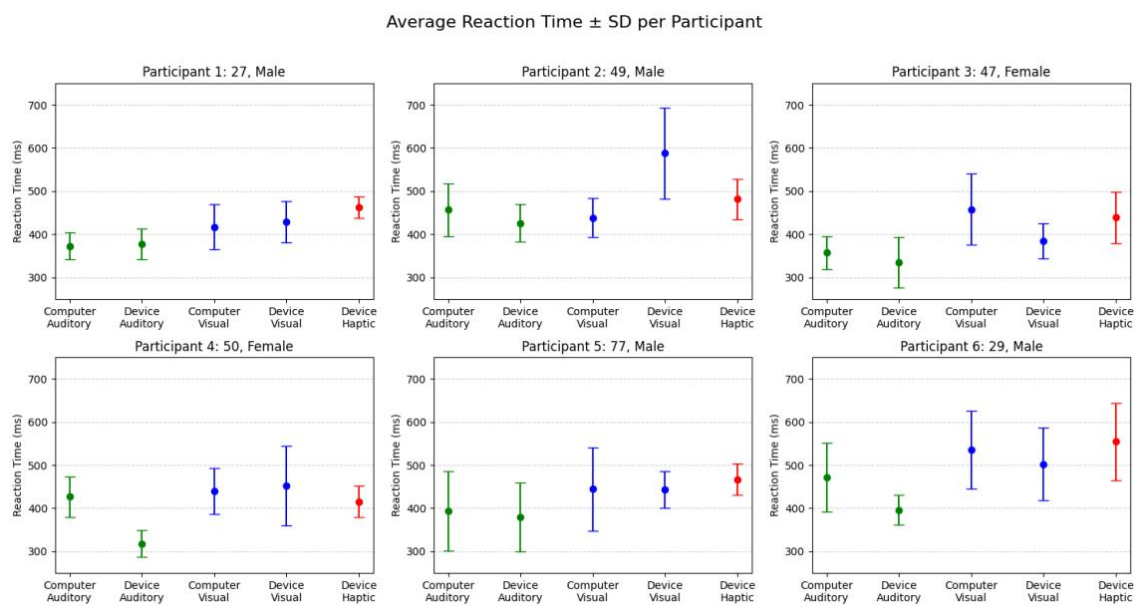

Fig. 10. Average reaction time ± standard deviation for each of the six participants across five testing conditions: computer-auditory, device-auditory, computer-visual, device-visual, and device-haptic. Each data point represents the mean reaction time for a specific participant and modality; error bars indicate ±1 standard deviation. Color coding reflects stimulus modality: green (auditory), blue (visual), and red (haptic). Trials with no recorded response or reaction times exceeding 1000 ms were excluded as outliers. A total of 4 out of 300 trials (1.33%) were removed

## 3.1 Auditory Reaction Time

Auditory stimuli yielded the fastest median reaction times across all participants, with device-auditory median RTs closely matching or outperforming median computer-auditory RTs in five of six participants. Participant 3 exhibited a 22.2 ms improvement on the device, while Participant 2 showed a greater 31 ms improvement. Participant 4 showed the largest discrepancy, responding 109 ms faster on the device-auditory test. While this result may reflect latency and variations in sensing, it is also possible that reduced attentiveness during the computer-based test contributed to the slower response. Given the low variability (SD < 50 ms) of RTs, the difference is unlikely to result from random fluctuations or system latency alone. Notably, the standard deviations (SDs) for auditory RTs were consistently small (e.g., SD < 50 ms for Participants 1 and 4), indicating high temporal precision and minimal system latency [48]. This performance reflects the direct neural routing of auditory stimuli and confirms the device's auditory interface is well-calibrated and reliable.

## 3.2 Visual Reaction Time

Median visual RTs were generally slower than auditory responses and displayed higher variability. This is probably because the input from the wearable device was a small LED, whereas the computer employed a large screen. In most participants (e.g., Participants 1, 2, 4 , and 5), device-visual RTs were longer and more variable than computer-visual RTs. This trend may be attributed to differences in visual contrast or luminance, and stimulus detectability in the wearable context [49]. Participant 3 and 6 were notable exceptions, with a device-visual RT 73 ms and  30 ms faster than the computer version. However, across all subjects, the larger standard deviations (up to ~100 ms) highlight a need for visual path optimization, particularly in timing control and screen responsiveness.

## 3.3 Haptic Reaction Time

The haptic modality, implemented only on the device, yielded reaction times between those of auditory and visual conditions. For most participants (e.g., Participants 1, 4, 5), haptic RTs clustered near 440–470 ms. Participant 6 showed the longest haptic RT (554.3 ms), accompanied by the largest SD (89.1 ms), suggesting sensitivity to actuator delay, skin contact variability, or tactile perception differences [50]. Despite variability, haptic RTs remained within established cognitive response windows [40]  and the modality offers strong potential for non-visual, passive RT monitoring, especially in inaccessible or sensory-overloaded environments.

## 3.4 Statistical Analysis

To quantitatively assess differences between modalities, paired t-tests were conducted on the reaction time means across six participants. Specifically, we compared computer- vs device-based auditory and visual responses, as well as device visual vs haptic modalities.

### 3.4.1 Paired Comparisons of Reaction Times

To further assess the device's validity across sensory modalities, paired t-tests were performed on individual participants comparing three key conditions: (1) Computer Auditory vs Device Auditory, (2) Computer Visual vs Device Visual, and (3) Device Visual vs Device Haptic. Each condition utilized n=10 trials per condition; If an outlier was excluded from a dataset, the corresponding value in the paired dataset was also excluded. As a result participant 3 (visual), participant 5 (auditory), participant 6 (auditory), and participant 6 (visual) each have n=9 trials per condition. The table below summarizes the t-statistics and p-values for each participant.

| Participant | Comparison | t-statistic | p-value |
|---|---|---|---|
| P1 | Computer Auditory vs Device Auditory | -0.40 | 0.695 |
| P1 | Computer Visual vs Device Visual | -0.45 | 0.665 |
| P1 | Device Visual vs Device Haptic | -2.139 | 0.061 |
| P2 | Computer Auditory vs Device Auditory | 1.22 | 0.253 |
| P2 | Computer Visual vs Device Visual | -4.20 | 0.002 |
| P2 | Device Visual vs Device Haptic | -2.23 | 0.053 |
| P3 | Computer Auditory vs Device Auditory | 0.90 | 0.390 |
| P3 | Computer Visual vs Device Visual | 2.14 | 0.065 |
| P3 | Device Visual vs Device Haptic | 0.72 | 0.489 |
| P4 | Computer Auditory vs Device Auditory | 4.59 | 0.001 |
| P4 | Computer Visual vs Device Visual | -0.31 | 0.763 |
| P4 | Device Visual vs Device Haptic | 1.03 | 0.330 |
| P5 | Computer Auditory vs Device Auditory | 0.53 | 0.611 |
| P5 | Computer Visual vs Device Visual | 0.05 | 0.958 |
| P5 | Device Visual vs Device Haptic | -0.65 | 0.529 |
| P6 | Computer Auditory vs Device Auditory | 2.77 | 0.024 |
| P6 | Computer Visual vs Device Visual | 0.12 | 0.908 |
| P6 | Device Visual vs Device Haptic | -0.43 | 0.673 |

Table 1: Individual participant statistic performance assessment.

The results of the paired t-tests revealed 3 statistically significant differences across conditions. For 4 participants, device-auditory responses did not significantly differ from computer-auditory responses, suggesting comparable auditory RT fidelity. However, in 2 participants (P4, P6), the computer auditory responses were slower than those measured by the device (p=0.001, p=0.024). Despite the statistical significance, the low standard deviations in device-based auditory RTs for both participants suggest that the differences may stem from individual participant variability rather than inconsistencies in device performance.

Visual modality comparisons showed no statistically significant differences for 5 participants. However, P2 demonstrated significantly longer and more variable device based reaction times compared to the computer (p=0.002). This statistical significance, paired with large variability in P2's RT (SD=105.7 ms) highlights a disparity in stimulus detectability across device and computer tests likely attributable to the computer's large display and the device's small LED.

The comparison between computer-visual and device-haptic responses showed no statistically significant differences, with p-values ranging from 0.053 to 0.673 across participants. This suggests that the haptic modality may be a viable substitute to visual stimuli for RT assessments in this device, particularly in natural scenarios where visual attention is limited or unavailable. However, some results near the significance threshold warrant further testing.

These findings support the utility of multimodal monitoring and suggest that the device provides differentiated, interpretable responses across sensory domains. Further testing in larger cohorts is warranted.

### 3.4.2 Lumped Paired Comparisons Across Participants

To complement the individual-level comparisons, we conducted paired t-tests by using the mean reaction times for each participant across conditions. With 6 participants, each test included n=6 paired samples. The table below summarizes the resulting t-statistics and p-values for each inter-participant comparison.

| Comparison | t-statistic | p-value |
|---|---|---|
| Computer Auditory vs Device Auditory | 2.36 | 0.065 |
| Computer Visual vs Device Visual | - 0.36 | 0.732 |
| Device Visual vs Device Haptic | -0.18 | 0.289 |

Table 2: Lumped performance statistics

The analysis revealed no statistically significant difference in lumped average reaction times between the computer-based and device-based auditory conditions (t=2.35 p=0.065). This supports the conclusion that the wrist-worn device can reliably and accurately measure auditory RT, but the result near the significance threshold indicates that further testing with larger samples should be done to confirm the effect. Additionally, no significant differences were observed for the visual modality (p = 0.732) or for the comparison between device visual and haptic modalities (p = 0.289), indicating comparable performance across these conditions when aggregated over participants. However, the high variability observed in visual responses suggests that additional investigation is needed into visual correlations. These results further support the device's cross-modality fidelity and highlight haptic monitoring as a particularly promising alternative in naturalistic scenarios where visual attention is limited.

## 3.5 Latency

As with all RT measuring systems, some latency is present in both the device and computer tests. Latency can mainly be attributed to two sources: (1) delivering stimulus and (2) detecting reactions.

### 3.5.1 Device Latency

Stimulus delivery latency is almost negligible for auditory and visual modalities as active buzzers and LEDs actuate within sub-millisecond ranges, well below human perceptual thresholds (<1 ms). Contrastingly, haptic feedback has a non-negligible hardware latency as a typical 3V, 12000 RPM coin vibration motor has a maximum rise time of 90 ms [44]. This is however a maximum, and the motor may reach a perceptible intensity before max speed is reached. Further research should be done to determine this threshold.

With respect to reaction detection, the Arduino's LSM9DS1 gyroscope operates at a fixed output data rate of 104 Hz [52], corresponding to a sampling interval of ~10 ms. This introduces a latency between 0 ms to 10 ms depending on when the reaction occurs within the gyroscope's sampling window.

### 3.5.2 Computer Latency

Computer-visual stimulus delivery latency arises from the screen's 60 Hz refresh rate [53], resulting in a screen update every ~17 ms. Similar to the gyroscope, this creates a latency window between 0 ms to 17 ms, depending on when the stimulus is applied relative to the screen's sampling cycle.

Auditory latency stems from both software buffering and hardware, and the computer's (2020 M1 macbook pro 13") speakers have been measured to add 14 ms of latency [54]

The computer uses its inbuilt keyboard to detect reactions. While specific latency data for the MacBook model used in this study is unavailable, most modern keyboards have a polling rate of 125 Hz [55], corresponding to a maximum latency of 8 ms which depends on when the reaction occurs in the polling cycle.

### 3.5.3 Overall Latency Analysis

Combining stimulus delivery and reaction detection delays, the estimated maximum total latencies are:

- Auditory: ~10 ms (device) vs. ~24 ms (computer)
- Visual: ~10 ms (device) vs. ~25 ms (computer)
- Haptic: up to ~100 ms (device)

It is important to note that these figures represent worst-case bounds, and real latencies depend on where the stimulus and reaction fall in the devices/computer's polling window. While the device demonstrates lower maximums in auditory and visual modalities by ~15 ms, this distinction remains small and unlikely to impact cross-platform comparisons. However, the haptic modality introduces substantially higher maximum latency due to motor rise time, which may affect comparisons with visual and auditory modalities. Further investigation is warranted into the absolute threshold of haptic stimulus with coin vibration motors, as users may perceive the stimulus before the motor reaches maximum speed.

### 3.4.3 Comparison with Prior Literature

Backyard Brains educational data [51] report ~170 ms for auditory, ~250 ms for visual, and ~150 ms for haptic RT. These RT times are shorter than those reported in our study. The difference may be due to the nature of our tests that attempted to simulate natural conditions and therefore distracted the participants. Nevertheless RT values observed in this study are in agreement with other reports. Yoshida, K. et al. [40] reported RTs ranging from $320 \pm 43$ ms (multimodal) to $528 \pm 105$ ms (visual-only on mobile devices) [43]. They found that tactile RTs were 28–34% faster than visual, and auditory RTs ~5% faster than visual. Our device's RTs (mostly 350–550 ms) fall within the ranges reported in these studies, validating both the physiological plausibility and practical utility of the new platform. Visual RTs in multimodal devices often exceed 500 ms, matching our observations of slower device visual responses. Haptic stimuli typically provide faster and more consistent responses, aligning with faster RTs in our haptic condition. These similarities validate that our device measures RT within expected human performance ranges for portable systems.

| Aspect | Sarkar & Rubinsky (2025) | Yoshida et al. (2023) | Cinaz et al. (2012) | Ivorra et al. (2008) |
|---|---|---|---|---|
| **Form Factor** | Wrist-worn, multimodal (visual, auditory, haptic) | Smartphone-based iOS app | Wrist-worn, haptic only | Wrist-worn haptic device |
| **Stimulus Types** | Visual (LED), auditory (buzzer), haptic (vibration motor) | Visual (screen flash), auditory (tone), haptic (vibration) | Haptic only (vibration motors) | Haptic only (vibration) |
| **Response Modality** | Natural wrist rotation measured by gyroscope | Button press on touchscreen | Wrist rotation via IMU | Wrist rotation via accelerometers |
| **Operating Environment** | Designed for naturalistic use during real-life activities | Controlled experiment while holding phone on table | Controlled dual-task protocol (idle/load) | Fully naturalistic 8-hour real-life deployment |
| **Participants** | 6 adults (ages 27–77), engaged in conversation during testing | 20 young adults (20–29 years old), in lab | 20 young adults in two experimental groups | 10 students, 8-hour naturalistic session |
| **Multimodal Comparison** | Full 3-modality validation (audio, visual, haptic) | 26 combinations of three modalities at two intensities | Not multimodal | Not multimodal |
| **Key Innovations** | First to demonstrate passive, real-world multi modal RT tracking via natural motion | Rich tri-modal analysis with intensity levels | First wearable go/no-go reaction time interface | First feasibility of minimally obtrusive wearable RT assessment |
| **Latency Consideration** | Quantified latency bounds by modality (e.g., motor rise time, polling) | Controlled but not hardware-calibrated | Fixed intervals, no detailed latency model | Basic delay accounted for, no latency benchmarking |
| **Statistical Rigor** | Individual and group-level t-tests; modality-by-modality comparison | ANOVA with interaction effects across levels and combinations | Repeated measures ANOVA, CV and error rates included | No formal stats; summary reporting |
| **Goal Orientation** | Ecological assessment for health and cognition | UX-focused insight into sensory interaction timing | Benchmarking wearable vs. desktop under dual-task load | Proof-of-concept of continuous unobtrusive RT tracking |
| **Technology Platform** | Arduino Nano BLE, 3-axis gyroscope | iPhone 11, Core Haptics, AVFoundation | ETHOS IMU + custom PCB | PIC microcontroller + dual accelerometers |

Table 3. Evolution of mobile RT technologies.

Table 3 compares the characteristics of four RT mobile technologies. The work in the study of this paper advances the field in four key ways:

1. Multimodal Stimulus Integration in a Naturalistic Context:

   For the first time, we combine auditory, visual, and haptic stimuli into a single wearable platform while using natural wrist rotation as the response—a method both reflexive and biomechanically congruent with daily activity. In contrast to the touchscreen presses used by Yoshida et al. or the structured button presses of desktop-based systems, our motion-based input preserves ecological validity without compromising signal fidelity.

2. Unobtrusive Real-World Deployment:

   Unlike previous studies that required fixed postures or direct user engagement, our design supports continuous use during naturalistic scenarios, including conversation and multitasking, without the need for gaze fixation or hand-eye coordination. This expands the potential for ambient cognitive monitoring, which neither smartphone apps nor lab-bound wearables have enabled effectively.

3. Cross-Modality Validation and Latency Profiling:

   We present detailed comparisons across five testing conditions (computer and device-based visual and auditory, and device-based haptic), enabling the first comprehensive modality-matched evaluation of a wearable system. Furthermore, we quantify stimulus-response latencies for each modality and platform, addressing a critical limitation in prior work where device latency was often unreported or uncontrolled.

4. Technological Maturity and Replicability:

   Our system uses open hardware (Arduino Nano BLE) and a documented stimulus logic pipeline that allows precise scheduling, synchronized data capture, and scalable testing. Prior wearable platforms were either proprietary (Cinaz et al.) or minimally described (Ivorra et al.), limiting replicability and downstream innovation. We also provide full documentation and source code, fostering transparency and adoption.

Together, these innovations position the present work as a culmination and convergence of previous approaches, combining modality-rich stimulus delivery, unobtrusive deployment, naturalistic interaction, and rigorous comparative validation in a single, low-cost, and extensible platform. This represents a substantial step forward for the use of RT as a practical biomarker in mobile cognitive health, neuroergonomics, fatigue monitoring, and ambient human-machine interface design.

4.Conclusion

This study introduces a novel, wrist-worn, multimodal reaction time (RT) monitoring system designed for passive, real-world cognitive assessment. By integrating auditory, visual, and haptic stimuli with natural wrist-rotation detection via a gyroscope, the device enables seamless, low-latency RT tracking during everyday activities—without requiring structured task engagement, gaze fixation, or user supervision. Validation experiments comparing the device to conventional computer-based RT assessments across five testing conditions demonstrated comparable performance and timing fidelity, particularly for auditory stimuli, which showed high precision. Haptic RTs emerged as a promising alternative to visual RT in environments where visual attention is limited, while visual RTs, though more variable, remained physiologically plausible.

Importantly, the system is built on an open-source, microcontroller-based architecture with transparent firmware, supporting reproducibility and customization by the broader research community. This platform is the first to combine passive operation, multimodal stimulus input, and naturalistic usability in a single, compact form factor. As a proof of concept, it establishes a strong foundation for future applications in mobile neurocognitive monitoring, fatigue detection, ambulatory screening for cognitive decline, and adaptive human–machine interface design. Future work will focus on extended-duration deployment, large-scale population studies, and integration with machine learning models for context-aware cognition tracking.

5. Ethical Considerations

This study involved six participants and was designed as a small-scale, end-to-end system test. The study was done under IRB Protocol ID 2020-08-13529 Haptic (non-verbal) communication. Informed consent was obtained from all participants, who were briefed on the study's purpose, procedures, and any potential risks. To protect participant privacy, all data collected was de-identified and securely stored in a database, in adherence to ethical standards for research with human subjects.

6. Additional data

Device Test Program (C++):
https://github.com/nsarkar7/reactionTimeDevice/blob/main/reactionTimeDevice.ino
Computer Test Program (HTML/JavaScript):
https://github.com/nsarkar7/reactionTimeDevice/blob/main/computerTest.html
Acknowledgement:
The authors used OpenAI's ChatGPT to assist in improving the grammar, clarity, and style of the manuscript. All content generated or suggested by the model was critically reviewed and independently verified by the authors. The authors take full responsibility for the integrity and accuracy of all statements presented in this work.